# Possible high-temperature magnetically topological material $Mn_3Bi_2Te_6$

Wen-Feng Wu[a,b], Han-Yu Wang[a,b], Wei-Hua Wang[c], Da-Yong Liu[d], Xiang-Long Yu[e,f], Zhi Zeng[a,b]\*, and Liang-Jian Zou[a,b]\*

a. Key Laboratory of Materials Physics, Institute of Solid State Physics, HFIPS, Chinese Academy of Sciences, Hefei 230031, China
b. Science Island Branch of Graduate School, University of Science and Technology of China, Hefei 230026, China
c. Department of Electronic Science and Engineering, and Tianjin Key Laboratory of Photo-Electronic Thin Film Device and Technology, Nankai University, Tianjin 300071, China
d. Department of Physics, School of Sciences, Nantong University, Nantong 226019, China
e. Shenzhen Institute for Quantum Science and Engineering, Southern University of Science and Technology, Shenzhen 518055, China
f. International Quantum Academy, Shenzhen 518048, China

## Abstract

The Mn-Bi-Te family displaying magnetism and non-trivial topological properties has received extensive attention. Here, we predict that the antiferromagnetic structure of $Mn_3Bi_2Te_6$ with three MnTe layers is energetically stable and the magnetic coupling strength of Mn-Mn is enhanced four times compared with that in the single MnTe layer of $MnBi_2Te_4$. The predicted Néel transition point is higher than 77 K, the liquid-nitrogen temperature. The topological properties show that with the variation of the MnTe layer from a single layer to three layers, the system transforms from a non-trivial topological phase to a trivial topological phase. Interestingly, the ferromagnetic state of $Mn_3Bi_2Te_6$ is a topological semimetal and it exhibits a topological transition from trivial to non-trivial induced by the magnetic transition. Our results enrich the Mn-Bi-Te family system, offer a new platform for studying topological phase transitions, and pave a new way to improve the working temperature of magnetically topological devices.

\* Correspondence author, zzeng@theory.issp.ac.cn; zou@theory.issp.ac.cn

# I. Introduction

Recent increasing attention has been paid to intrinsic magnetic topological materials due to their unusual properties and promising applications in low-power consumption devices [1-27]. As the first material intrinsically combining magnetism with topology, MnBi$_2$Te$_4$ (MBT) was found to be antiferromagnetic (AFM) [9–12,14,28]. Meanwhile, experimental and theoretical studies showed that it poses the quantum anomalous Hall effect (QAHE) and the topological magnetoelectric effect (TME) [10-12,14,18]. For its bulk phase, MBT appears as a topologically non-trivial surface state on a surface with joint symmetry $S = \mathcal{T}\tau$, forming the so-called magnetic topological insulator (MTI) [6,7,9,28]. For MBT thin films, the peculiar phenomenon is demonstrated as *Chern* insulators in odd-number layers and as axion insulators in even-number layers [14,16,17,29,30]. Previous researches demonstrate that the AFM transition temperature ($T_N$) of MBT is ~25 K [6,7,9,10,13], which is too low to be widely applied in magnetic or spintronic devices at room temperature. As it is known, there are several ways to improve $T_N$. Firstly, the spatial distance between magnetic layers is reduced to enhance the interlayer magnetic couplings and suppress the spin fluctuations. Secondly, the spin-orbit coupling (SOC) is increased to enhance magnetic anisotropy. In addition, the number of magnetic layers is increased to enhance the magnetic interaction between Mn spins. These ansatzes provide potential methods to lift the Néel temperature $T_N$ or the Curie temperature $T_C$.

With one of these thought lines, an effort has been made to search for more Mn-Bi-Te systems. Zhang *et al.* experimentally synthesized two MnTe layers compound Mn$_2$Bi$_2$Te$_5$ (M$_2$BT) [3], and the primary theoretical study suggested that M$_2$BT is an intrinsic AFM axion insulator displaying TME and chiral magnetic effect (CME) [1,3,4,19]. Theoretical studies suggest that due to the small exchange energy, M$_2$BT can easily transit from a topologically-trivial AFM phase to a topological Weyl semimetal when an external field is applied to lift the magnetic transition temperature [4]. Since M$_2$BT has two interacting MnTe layers in a unit cell, it should have stronger magnetic coupling and higher $T_N$. However, experimental measurement on magnetic susceptibility showed that the Néel temperature $T_N$ of M$_2$BT was about 24 K [1,3], almost identical to that of MBT. This count-intuitive result deserves further study.

In this paper, we have explored the thermodynamic stability, electronic structures, magnetic and topological properties of tri-MnTe-layer compound Mn$_3$Bi$_2$Te$_6$ (M$_3$BT) by using the first-principles calculations to design novel magnetic topological materials with much higher magnetic critical temperature, or even up to room temperature. We find that the magnetic coupling strength of Mn-Mn in M$_3$BT increases to four times that in a single MnTe layer of MBT. Thus, the predicted Néel transition point is improved above the liquid-nitrogen temperature (77 K). Moreover, a rare phenomenon of topological phase transition induced by magnetic phase transition is found in M$_3$BT. We organize the rest of this paper as follows: in section II, we show the crystal structure of M$_3$BT and discuss the structural stability. In section III, we investigate the magnetic properties of M$_3$BT, and the result shows that the exchange energy significantly enhances in comparison with the single MnTe layer of MBT. In section IV, we explore the topological properties and topological phase transitions in M$_3$BT. The final section is devoted to a brief discussion and summary.

## II. Structural properties of M$_3$BT

As mentioned above, the experimental results show that the antiferromagnetic transition temperature of MBT is ~ 24 K, and relevant theoretical calculations show that the energy difference between the antiferromagnetic structure and ferromagnetic structure ($\Delta E_{A/F}$) is -2.8 meV in bulk MBT [14]. Such a small exchange energy is attributed to very weak magnetic coupling and thus low T$_N$ due to the van der Waals interaction between the MnTe layer and the Bi$_2$Te$_3$ layer. To improve T$_N$, one effective strategy is to appropriately increase the thickness of the MnTe layer. In previous reports, M$_2$BT is synthesized and it possesses a layered rhombohedra crystal structure with the space group of $P\bar{3}m1$ (No.164); a = b = 4.288 Å, c = 17.150 Å) with the NaCl-type (ABC) stacking of nine-atom layers, known as a nonuple layer (NL) and there is a van der Waals interaction between different NLs [1,3]. We find that the exchange energy $\Delta E_{A/F}$ = -7.6 meV after full structural relaxation with the structural parameters mentioned above, which is very close to the results reported in the literature [3]. To further improve the Mn-Mn magnetic coupling strength, we increase the number of MnTe layers, predicting a stable material M$_3$BT with three MnTe layers. The computational details are presented in Appendix.

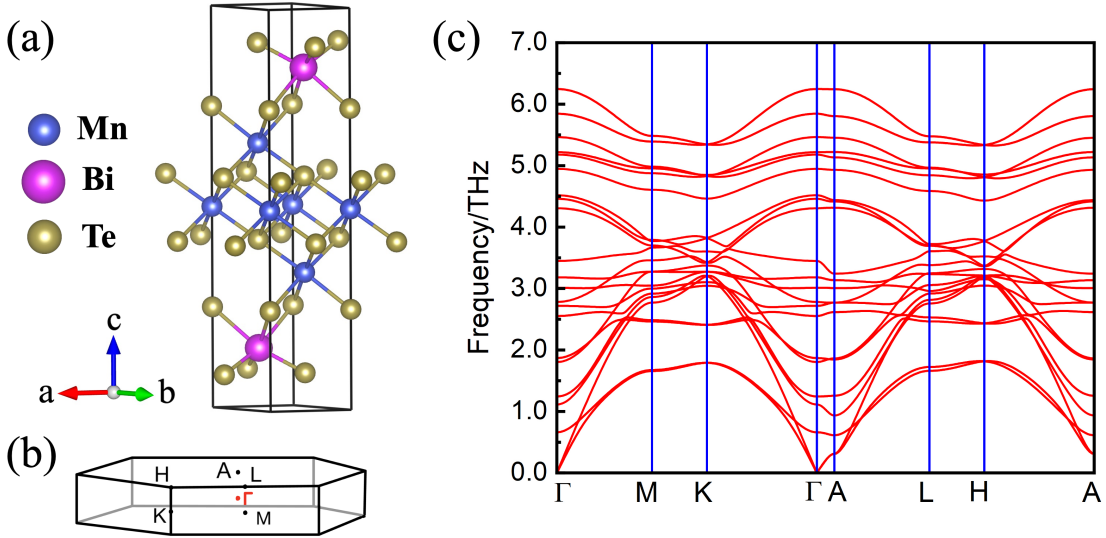

FIG. 1 Crystal structure and stability of M$_3$BT. (a) The crystal structure of M$_3$BT cells. Blue, red, and yellow balls are Mn, Bi and Te atoms, respectively. (b) The reduced Brillouin zone of M$_3$BT, where high symmetric points are marked. (c) Phonon dispersion spectra of M$_3$BT.

The crystal structure of M$_3$BT is shown in FIG.1 (a), which is a unit cell with undecuple layers (ULs) of atoms. Similar to MBT and M$_2$BT, the middle of the structural unit is three magnetic intercalated MnTe layers, and the two ends are the topological layer Bi$_2$Te$_3$. After fully relaxing the structure, the lattice constants of the M$_3$BT are a = b = 4.100 Å and c = 19.300 Å, respectively. It has the space group $P\bar{3}m1$ (No.164) with inversion symmetry if the spin moments of Mn are ignored. Subsequently, the finite-displacement method is used to calculate the phonon spectrum of M$_3$BT (FIG.1b). The absence of imaginal frequencies in the phonon dispersions indicates that predicted M$_3$BT is dynamically stable.

### III. Magnetic properties of M$_3$BT

Since the 3$d$ electrons of Mn atoms possess magnetic moments, the effect of magnetic interaction between Mn atoms should be considered. By doubling the structural unit along the $z$-axis, we found that the ferrimagnetic (FI) state within the supercell and the magnetic antiferromagnetic (AFM) state between adjacent ULs (FIG. 2a) would be preferable, with the energy difference $\Delta E_{A/F}$ ($E$(FI-AFM) − $E$(FM)) = -11.7 (meV / Mn pair). It implies that the FM order is not favored. Nevertheless, for each in-plane Mn layer, the two nearest Mn are connected by Te atoms and the bond angle is about 95 degrees, the superexchange interaction will cause FM order in the in-plane Mn layer according to the Goodenough-Kanamori rules. Two other possible configurations, FM-AFM and FI-FM (FIG. 2 (b)-(c)) were considered and their energies were higher than those of FI-AFM. Thus, the FI-AFM is the most stable magnetic configuration in $M_3BT$.

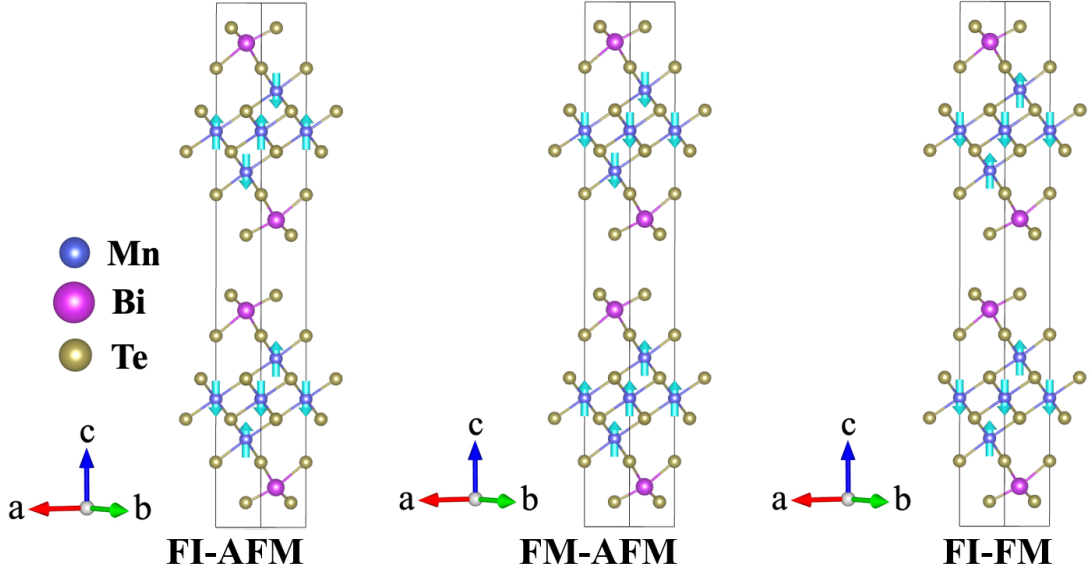

FIG. 2 Possible non-FM magnetic configurations of $M_3BT$. (a) FI-AFM configuration which exhibits the characteristics of AFM. Other configurations in (b) and (c) are not stable.

It is important to note that the Coulomb interactions play crucial roles in the present correlated topological systems, the ground-state magnetic configurations of these insulating compounds obtained by the first-principles calculations strongly depend on the strength of U. The above result is obtained with the correlation strength of U = 3.0 eV for Mn-3$d$ electrons according to previous theoretical research of $M_2BT$ [2,3]. Considering the intrinsically strong correlation of Mn-3$d$ electrons, we present the exchange energy difference with the FI-AFM state and the FM state of $M_3BT$ when U varies from 3.0 to 6.0 eV, as listed in Table 1. We find that the FM state is the most stable magnetic phase when U = 4.0 - 6.0 eV, and the ground-state magnetic configuration of $M_3BT$ strongly depends on the Coulomb interaction strength. Therefore, further experiments are expected to determine the realistic magnetic configuration of $M_3BT$.

Table 1. The relative energy of two possible magnetic configurations at different U values. $\Delta E_{A/F}$ is the energy difference between the FI-AFM state and the FM state of $M_3BT$, reflecting the magnetic exchange interaction between the Mn spin pair.

| U [eV] | 3 | 4 | 5 | 6 |
|---|---|---|---|---|
| $\Delta E_{A/F}$ [eV / (Mn pair)] | -0.0117 | 0.0158 | 0.0265 | 0.0363 |

The dependence of the magnetic coupling strength between Mn-Mn spins on different MnTe layers is shown in Table 2. Here, the theoretical result of MBT in the literature [14] is cited for comparison. The coupling strength becomes strong with an increasing number of MnTe layers. In detail, the spin coupling strength of $M_2BT$ is about threefold of MBT and the strength of $M_3BT$ is about four times that of MBT. On the one hand, it stems from the enhancement of superexchange interaction within the interlayer Mn-Mn spins as the number of MnTe layers increases. On the other hand, the increase of the nearest-neighboring number suppresses the quantum fluctuations of Mn spins, which results in the enhanced coupling strength of Mn-$3d$ electrons and hence considerably increases of magnetic transition temperatures of $M_2BT$ and $M_3BT$.

Table 2. Preferable configuration (marked in blue) and magnetic exchange energy of $Mn_nBi_2Te_{3+n}$ (n=1,2,3).

| Materials | Magnetic configuration | $\Delta E_{A/F}$ [eV / (Mn pair)] |
|---|---|---|
| $MnBi_2Te_4$ | FM | -0.0028 |
| | AFM | |
| $Mn_2Bi_2Te_5$ | FM | -0.0076 |
| | AFM | |
| $Mn_3Bi_2Te_6$ | FM | -0.0117 |
| | FI-AFM | |

In addition, previous research found that the small magnetic coupling in MBT can modulate the magnetic configurations by the external field [12,13,20,30]. Similarly, $M_2BT$ can also undergo the same process [4]. For the $M_3BT$, since the coupling strength on the Mn pair of the FI-AFM configuration is 11.7 meV lower than that of FM, we predict that the external magnetic field can also induce the AFM-FM transformation. Therefore, it is believed that FI-AFM can be driven into FM by the magnetic field. Further analysis shows that the FM phase is a topological Weyl semi-metal. During the magnetic phase transition, $M_3BT$ would undergo the transition from insulator to metal caused by the magnetic transition.

## IV. Topological properties in $M_3BT$

The previous work showed that MBT is an AFM topological insulator and it has a band gap of 77 meV [18] since the topological surface state wavefunctions extend to MnTe and hybridize with Mn $3d$-electrons [6–8,18,28]. For AFM $M_2BT$, the system is a dynamic axion insulator, which still has topological properties [3]. Once these two materials enter the FM state, they behaver the Weyl semi-metals [4,12]. While for $M_3BT$ in the magnetic ground state, we calculated the Wannier charge center (WCC) (FIG. A1), which can be employed to determine the topology of the FI-AFM $M_3BT$.

The system is in a trivial state and the detailed band structure analysis shows no band inversion occurrence.

The above results show that from $M_2BT$ to $M_3BT$, the system undergoes a topological phase transition driven by the number of MnTe layers. From the projected density of states (PDOS), we find that the reason for $M_3BT$ being topologically trivial is that its $d$ electrons are extended to the vicinity of the Fermi surface, which destroy the original surface states with the contributed from Te $p$ electrons and the system loses its topology. We suggest that this "magnetic extension" is excessive. This situation is similar to over-doped Cr in $Bi_2Se_3$ [31], where the lighter element has a weak spin-orbit coupling. With the increase of Cr concentration, the Curie temperature is increased. However, the $p$-$d$ hybridization leads to the increase of the $d$ orbital component of the band near the Fermi surface, which decreases the spin-orbit coupling and destroys the system's topological property to a certain extent.

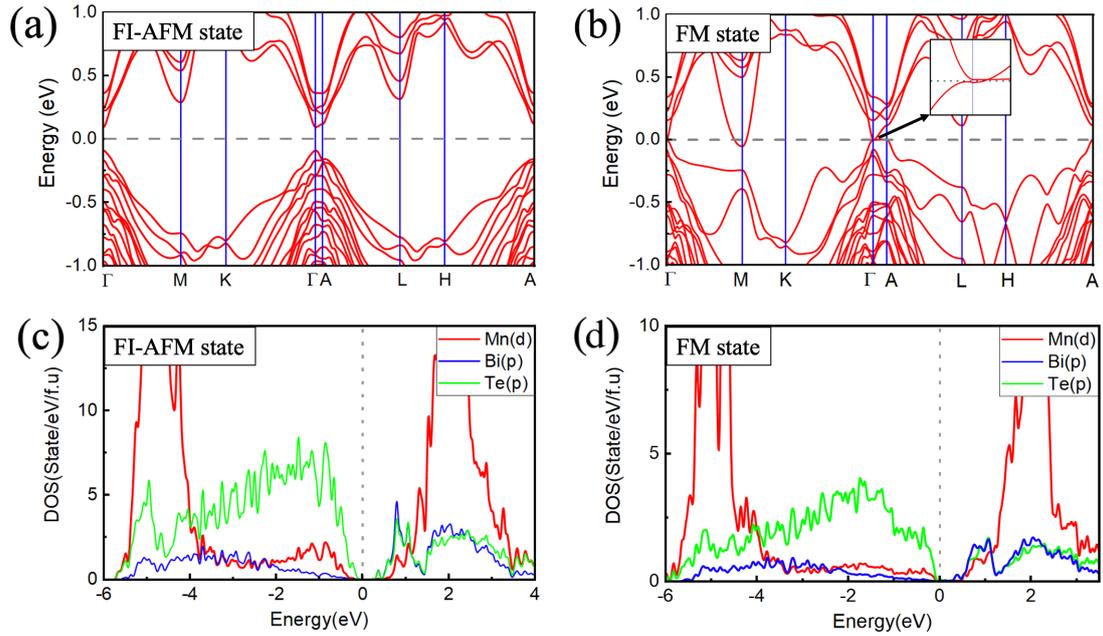

FIG 3. The electronic structures and projected density of states (PDOS) of $M_3BT$. (a) Band structure for FI-AFM state. (b) Band structure for FM state. The band dispersions near the gamma point have been zoomed in. (c)-(d) The PDOS for the FI-AFM and the FM states of $M_3BT$. The red, blue, and green lines represent the $d$ orbitals of Mn, the $p$ orbitals of Bi, and the $p$ orbitals of Te, respectively.

Interestingly, in the FM state, the situations are very different. The electronic structures in FIG. 3 (a) shows that FI-AFM $M_3BT$ is an insulator with a direct band gap of 183 meV. But in the FM state, an electronic pocket appears at point M in the band structures of $M_3BT$ (FIG. 3 (b)). The band crossover appears in the path from $\Gamma$ to A, indicating that it is a Weyl semi-metal. The FM phase's magnetic space group is denoted as 164.89 in the Belov-Neronova-Smirnova (BNS) notation [32]. Similar to Liu *et al.*'s work [33], according to the existing 1651 magnetic space groups, we use the symmetry operations, the small core presentations, and the effective $k \cdot p$ Hamiltonians to systematically classify the emergent particles in magnetic materials and find the topological properties of materials. The unitary subgroup for the FM state of $M_3BT$ is No.147, and it can be judged that it carries the unit chiral topological charge [33]. As mentioned above, since the energy difference between the two magnetic configurations FI-AFM and FM of $M_3BT$ is not large, an external magnetic field could modulate the magnetic configurations. Therefore, our theoretical calculations predict that $M_3BT$ can transform from the topologically trivial FI-AFM insulating state to the non-trivial topological FM Weyl semi-metallic state by applying an external magnetic field.

## V. Discussion and conclusion

We have shown that it is an effective way to lift the magnetic transition temperature of topological magnetic materials by increasing the number of MnTe layers. We notice that besides recent Zhang *et al.*'s work [3], Tang *et al.* constructed and studied the $XYBi_2Te_5$ (X, Y =Mn, Ni, V, Eu) series system [34], and they showed the magnetic atom layers in these materials enhance the magnetic ordering temperature while keeping the topologically nontrivial properties They suggested that the reason why the magnetic transition temperature of $Mn_2Bi_2Te_5$ does not rise considerably is that the superexchange interactions of Mn atoms via the 90 degrees path and the 180 degrees path cancel out each other. However, the literature [35] showed through a strict calculation that in a similar structure of $d^5$ compounds, the magnetic exchange integral J of the two paths cannot cancel out with each other.

Our work has predicted a new Mn-Bi-Te family material $M_3BT$ with three layers of MnTe and its magnetic ground state exhibits the AFM configuration. With the increase of MnTe layers in Mn-Bi-Te family, the calculated magnetic coupling strength between Mn spins also increases, indicating that $M_2BT$ and $M_3BT$ should have higher $T_N$ relative to MBT. By analyzing the topological properties, we predict that the ground state of $M_3BT$ is a topologically trivial antiferromagnetic insulator, demonstrating that with the increase of the number of MnTe layers, the system undergoes a transition from topologically non-trivial to topologically trivial characterization. On the other hand, by an applied magnetic field, $M_3BT$ becomes an FM Weyl semi-metal with non-trivial topology, displaying that $M_3BT$ can undergo a topological transition from trivial to non-trivial induced by the magnetic transition, accompanied by the insulating-metal transition. Our results enrich the Mn-Bi-Te family system, provide theoretical predictions and suggestions for exploring the magnetic properties of the Mn-Bi-Te family, and shed light on designing and exploring the topological phase transitions in magnetic topological materials.

## Acknowledgments

The authors thank Dr. Fan Wei for his helpful suggestions on the discussions of magnetism in this work. The authors thank the support from the NSFC of China under Grant nos. 11974354, 11774350 and 51727806. X.L.Y. acknowledges the support by the Natural Science Foundation of Guangdong Province (Grant No. 2023A1515011852). Numerical calculations were performed at the Center for Computational Science of CASHIPS, the ScGrid of the Supercomputing Center, the Computer Network Information Center of CAS, and the Hefei Advanced Computing Center.

# Appendix

## A. Computational details of first-principles calculations

Our first-principles calculations were made within Vienna *ab initio* Simulation Package (VASP) framework [36,37]. The exchange relation between electrons and ions was obtained by using the Perdew-Burke-Ernzerhof (PBE) functional of the generalized gradient correction approximation (GGA) [38]. The cut-off energy for plane wave expansion was set to 410 eV. The energy and force convergence thresholds were set to $10^{-8}$ eV and 0.01 eV/Å. The Brillouin zone integral adopts the Monkhorst-Pack method, and the k-point grid was set as 11 × 11 × 3. The DFT-D3 method is used to include the van der Waals correction. Considering the correlation effect of the 3$d$ orbitals of Mn, U = 3 eV is set according to the relevant works of literature. Spin-orbit coupling (SOC) effect is included in self-consistent calculations. WANNIER90 code is adopted to build wannier functions and Mn-$d$, Bi-$p$, and Te-$p$ states are treated as the initial projections according to the orbital weight on the fermi level of Mn-Bi-Te family materials [39,40]. The wannier charge center (WCC) for topological properties is calculated by the iterative Green's function implemented in WANNIERTOOLS [41].

## B. The topological properties of M$_3$BT in the FI-AFM state

According to Zak's theory, we can transform Bloch representation into Wannier representation so that one can judge whether a system has topological properties from WCC. For the state of FI-AFM, the pattern of WCC shown in FIG. A1 indicates that the system is in a trivial state.

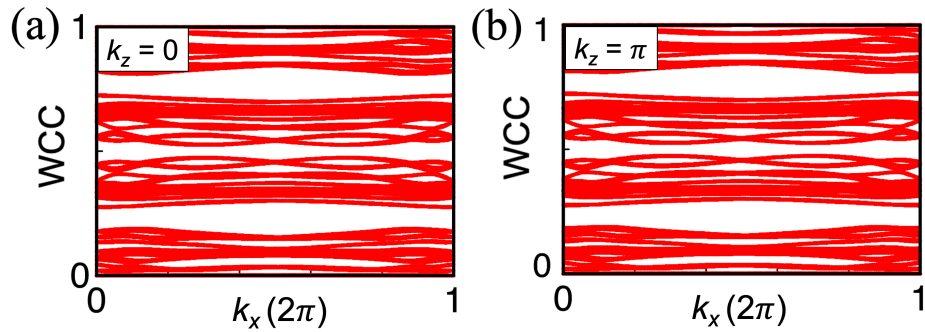

FIG. A1 The WCC of M$_3$BT in the FI-AFM state.

# References

[1] L. Cao, S. Han, Y.-Y. Lv, D. Wang, Y.-C. Luo, Y.-Y. Zhang, S.-H. Yao, J. Zhou, Y. B. Chen, H. Zhang and Y.-F. Chen, Growth and Characterization of the Dynamical Axion Insulator Candidate $Mn_2Bi_2Te_5$ with Intrinsic Antiferromagnetism, Phys. Rev. B **104**, 054421 (2021).

[2] Y. Li, Y. Jiang, J. Zhang, Z. Liu, Z. Yang, and J. Wang, Intrinsic Topological Phases in $Mn_2Bi_2Te_5$ Tuned by the Layer Magnetization, Phys. Rev. B **102**, 121107 (2020).

[3] J. Zhang, D. Wang, M. Shi, T. Zhu, H. Zhang, and J. Wang, Large Dynamical Axion Field in Topological Antiferromagnetic Insulator $Mn_2Bi_2Te_5$, Chin. Phys. Lett. **37**, 077304 (2020).

[4] S. V. Eremeev, M. M. Otrokov, A. Ernst, and E. V. Chulkov, Magnetic Ordering and Topology in $Mn_2Bi_2Te_5$ van Der Waals Materials, Phys. Rev. B **105**, 195105 (2022).

[5] Y. Ren, S. Ke, W.-K. Lou, and K. Chang, Quantum Phase Transitions Driven by Sliding in Bilayer $MnBi_2Te_4$, Phys. Rev. B **106**, 235302 (2022).

[6] J.-Q. Yan, Q. Zhang, T. Heitmann, Z. Huang, K. Y. Chen, J.-G. Cheng, W. Wu, D. Vaknin, B. C. Sales, and R. J. McQueeney, Crystal Growth and Magnetic Structure of $MnBi_2Te_4$, Phys. Rev. Mater. **3**, 064202 (2019).

[7] Y. Yuan, X. Wang, H. Li, J. Li, Y. Ji, Z. Hao, Y. Wu, K. He, Y. Wang, Y. Xu, W. Duan, W. Li and Q.-K. Xue, Electronic States and Magnetic Response of $MnBi_2Te_4$ by Scanning Tunneling Microscopy and Spectroscopy, Nano Lett. 7 (2020).

[8] P. Swatek, Y. Wu, L.-L. Wang, K. Lee, B. Schrunk, J. Yan, and A. Kaminski, Gapless Dirac Surface States in the Antiferromagnetic Topological Insulator $MnBi_2Te_4$, Phys. Rev. B **101**, 161109 (2020).

[9] M. M. Otrokov, I. I. Klimovskikh, H. Bentmann, D. Estyunin, A. Zeugner, Z. S. Aliev, S. Gaß, A. U. B. Wolter, A. V. Koroleva, A. M. Shikin, M. Blanco-Rey, M. Hoffmann, I. P. Rusinov, A. Y. Vyazovskaya, S. V. Eremeev, Y. M. Koroteev, V. M. Kuznetsov, F. Freyse, J. Sánchez-Barriga, I. R. Amiraslanov, M. B. Babanly, N. T. Mamedov, N. A. Abdullayev, V. N. Zverev, A. Alfonsov, V. Kataev, B. Büchner, E. F. Schwier, S. Kumar, A. Kimura, L. Petaccia, G. Di Santo, R. C. Vidal, S. Schatz, K. Kißner, M. Ünzelmann, C. H. Min, S. Moser, T. R. F. Peixoto, F. Reinert, A. Ernst, P. M. Echenique, A. Isaeva and E. V. Chulkov, Prediction and Observation of an Antiferromagnetic Topological Insulator, Nature **576**, 7787 (2019).

[10] Y. Deng, Y. Yu, M. Z. Shi, Z. Guo, Z. Xu, J. Wang, X. H. Chen, and Y. Zhang, Quantum Anomalous Hall Effect in Intrinsic Magnetic Topological Insulator $MnBi_2Te_4$, Science **367**, 895 (2020).

[11] S. H. Lee, Y. Zhu, Y. Wang, L. Miao, T. Pillsbury, H. Yi, S. Kempinger, J. Hu, C. A. Heikes, P. Quarterman, W. Ratcliff, J. A. Borchers, H. Zhang, X. Ke, D. Graf, N. Alem, C.-Z. Chang, N. Samarth and Z. Mao, Spin Scattering and Noncollinear Spin Structure-Induced Intrinsic Anomalous Hall Effect in Antiferromagnetic Topological Insulator $MnBi_2Te_4$, Phys. Rev. Res. **1**, 012011 (2019).

[12] D. Zhang, M. Shi, T. Zhu, D. Xing, H. Zhang, and J. Wang, Topological Axion States in the Magnetic Insulator $MnBi_2Te_4$ with the Quantized Magnetoelectric Effect, Phys. Rev. Lett. **122**, 206401 (2019).

[13] Z. Li, J. Li, K. He, X. Wan, W. Duan, and Y. Xu, Tunable Interlayer Magnetism and Band Topology in van Der Waals Heterostructures of $MnBi_2Te_4$-Family Materials, Phys. Rev. B **102**,

081107 (2020).

[14] M. M. Otrokov, I. P. Rusinov, M. Blanco-Rey, M. Hoffmann, A. Yu. Vyazovskaya, S. V. Eremeev, A. Ernst, P. M. Echenique, A. Arnau, and E. V. Chulkov, Unique Thickness-Dependent Properties of the van Der Waals Interlayer Antiferromagnet $MnBi_2Te_4$ Films, Phys. Rev. Lett. **122**, 107202 (2019).

[15] R. Li, J. Wang, X.-L. Qi, and S.-C. Zhang, Dynamical Axion Field in Topological Magnetic Insulators, Nat. Phys. **6**, 284 (2010).

[16] H. Wang, D. Wang, Z. Yang, M. Shi, J. Ruan, D. Xing, J. Wang, and H. Zhang, Dynamical Axion State with Hidden Pseudospin Chern Numbers in $MnBi_2Te_4$-Based Heterostructures, Phys. Rev. B **101**, 081109 (2020).

[17] S.-H. Su, J.-T. Chang, P.-Y. Chuang, M.-C. Tsai, Y.-W. Peng, M. K. Lee, C.-M. Cheng, and J.-C. A. Huang, Epitaxial Growth and Structural Characterizations of $MnBi_2Te_4$ Thin Films in Nanoscale, Nanomaterials **11**, 12 (2021).

[18] M. M. Otrokov, T. V. Menshchikova, M. G. Vergniory, I. P. Rusinov, A. Y. Vyazovskaya, Y. M. Koroteev, G. Bihlmayer, A. Ernst, P. M. Echenique, A. Arnau and E. V. Chulkov, Highly-Ordered Wide Bandgap Materials for Quantized Anomalous Hall and Magnetoelectric Effects, 2D Mater. **4**, 025082 (2017).

[19] M. Wang, H. Liu, and X. C. Xie, New Type of Anticommutative Dynamical Magnetoelectric Response, Phys. Rev. Lett. **128**, 236601 (2022).

[20] P. Chen, Q. Yao, J. Xu, Q. Sun, A. J. Grutter, P. Quarterman, P. P. Balakrishnan, C. J. Kinane, A. J. Caruana, S. Langridge, A. Li, B. Achinuq, E. Heppell, Y. Ji, S. Liu, B. Cui, J. Liu, P. Huang, Z. Liu, G. Yu, F. Xiu, T. Hesjedal, J. Zou, X. Han, H. Zhang, Y. Yang and X. Kou, Tailoring the magnetic exchange interaction in $MnBi_2Te_4$ superlattices via the intercalation of ferromagnetic layers, Nat. Electron. **6**, 18–27 (2023)

[21] H. Wang, N. Mao, C. Niu, S. Shen, M.-H. Whangbo, B. Huang, and Y. Dai, Ferromagnetic Dual Topological Insulator in a Two-Dimensional Honeycomb Lattice, Mater. Horiz. **7**, 2431 (2020).

[22] B. Huang, G. Clark, E. Navarro-Moratalla, D. R. Klein, R. Cheng, K. L. Seyler, D. Zhong, E. Schmidgall, M. A. McGuire, D. H. Cobden, W. Yao, D. Xiao, P. Jarillo-Herrero and X. Xu, Layer-Dependent Ferromagnetism in a van Der Waals Crystal down to the Monolayer Limit, Nature **546**, 270 (2017).

[23] X. Hu, N. Mao, H. Wang, Y. Dai, B. Huang, and C. Niu, Quantum Spin Hall Effect in Antiferromagnetic Topological Heterobilayers, Phys. Rev. B **103**, 085109 (2021).

[24] Z.-X. Li, Y. Cao, and P. Yan, Topological Insulators and Semimetals in Classical Magnetic Systems, Phys. Rep. **915**, 1 (2021).

[25] F. Zhu, L. Zhang, X. Wang, F. J. dos Santos, J. Song, T. Mueller, K. Schmalzl, W. F. Schmidt, A. Ivanov, J. T. Park, J. Xu, J. Ma, S. Lounis, S. Blügel, Y. Mokrousov, Y. Su and T. Brückel, Topological Magnon Insulators in Two-Dimensional van Der Waals Ferromagnets $CrSiTe_3$ and $CrGeTe_3$: Toward Intrinsic Gap-Tunability, Sci. Adv. **7**, eabi7532 (2021).

[26] S. Jia, S.-Y. Xu, and M. Z. Hasan, Weyl Semimetals, Fermi Arcs and Chiral Anomalies, Nat. Mater. **15**, 1140 (2016).

[27] Y. Tokura, K. Yasuda, and A. Tsukazaki, Magnetic Topological Insulators, Nat. Rev. Phys. **1**, 126 (2019).

[28] J. G. Yan Gong and J. G. Yan Gong, Experimental Realization of an Intrinsic Magnetic

Topological Insulator, Chin. Phys. Lett. **36**, 76801 (2019).

[29] J. Choe, D. Lujan, M. Rodriguez-Vega, Z. Ye, A. Leonardo, J. Quan, T. N. Nunley, L.-J. Chang, S.-F. Lee, J. Yan, G. A. Fiete, R. He and X. Li, Electron-Phonon and Spin-Lattice Coupling in Atomically Thin Layers of $MnBi_2Te_4$, Nano Lett. **21**, 6139 (2021).

[30] Z. Li, J. Li, K. He, X. Wan, W. Duan, and Y. Xu, Tunable Interlayer Magnetism and Band Topology in van Der Waals Heterostructures of $MnBi_2Te_4$-Family Materials, Phys. Rev. B **102**, 081107 (2020).

[31] Y. Ou, C. Liu, L. Zhang, Y. Feng, G. Jiang, D. Zhao, Y. Zang, Q. Zhang, L. Gu, Y. Wang, K. He, X. Ma and Q.-K. Xue, Heavily Cr-Doped $(Bi,Sb)_2Te_3$ as a Ferromagnetic Insulator with Electrically Tunable Conductivity, APL Mater. **4**, 086101 (2016).

[32] B. N. V, The 1651 Shubnikov Groups, Kristallografiya **2**, 315 (1957).

[33] G.-B. Liu, Z. Zhang, Z.-M. Yu, S. A. Yang, and Y. Yao, Systematic Investigation of Emergent Particles in Type-III Magnetic Space Groups, Phys. Rev. B **105**, 085117 (2022).

[34] X.-Y. Tang, Z. Li, F. Xue, P. Ji, Z. Zhang, X. Feng, Y. Xu, Q. Wu, and K. He, Intrinsic High-Temperature Quantum Anomalous Hall Effect and Tunable Magnetic Topological Phases in $XYBi_2Te_5$, arXiv:2211.08163.

[35] H. Liu and G. Khaliullin, Pseudospin Exchange Interactions in $d^7$ Cobalt Compounds: Possible Realization of the Kitaev Model, Phys. Rev. B **97**, 014407 (2018).

[36] G. Kresse and J. Hafner, Ab Initio Molecular-Dynamics Simulation of the Liquid-Metal--Amorphous-Semiconductor Transition in Germanium, Phys. Rev. B **49**, 14251 (1994).

[37] G. Kresse and J. Furthmüller, Efficient Iterative Schemes for Ab Initio Total-Energy Calculations Using a Plane-Wave Basis Set, Phys. Rev. B **54**, 11169 (1996).

[38] J. P. Perdew, K. Burke, and M. Ernzerhof, Generalized Gradient Approximation Made Simple, Phys. Rev. Lett. **77**, 3865 (1996).

[39] N. Marzari, A. A. Mostofi, J. R. Yates, I. Souza, and D. Vanderbilt, Maximally Localized Wannier Functions: Theory and Applications, Rev. Mod. Phys. **84**, 1419 (2012).

[40] G. Pizzi, V. Vitale, R. Arita, S. Blügel, F. Freimuth, G. Géranton, M. Gibertini, D. Gresch, C. Johnson, T. Koretsune, J. Ibañez-Azpiroz, H. Lee, J.-M. Lihm, D. Marchand, A. Marrazzo, Y. Mokrousov, J. I. Mustafa, Y. Nohara, Y. Nomura, L. Paulatto, S. Poncé, T. Ponweiser, J. Qiao, F. Thöle, S. S. Tsirkin, M. Wierzbowska, N. Marzari, D. Vanderbilt, I. Souza, A. A. Mostofi and J. R. Yates, Wannier90 as a Community Code: New Features and Applications, J. Phys. Condens. Matter **32**, 165902 (2020).

[41] Q. Wu, S. Zhang, H.-F. Song, M. Troyer, and A. A. Soluyanov, WannierTools: An open-source software package for novel topological materials, Computer Physics Communications, **224**, 405 (2018).